\documentclass[%
 aip,
 amsmath,amssymb,
 reprint,%
]{revtex4-1}

\usepackage{graphicx}
\usepackage{dcolumn}
\usepackage{bm}

\usepackage[utf8]{inputenc}
\usepackage[T1]{fontenc}

\usepackage{amssymb}
\usepackage{amsbsy}
\usepackage{color}
\usepackage{amscd}
\usepackage{mathrsfs}
\usepackage{gensymb}

\newcommand{\RR}[1]{{\color{black}{{#1}}}}

\usepackage{mathrsfs}
\usepackage{calligra}
\DeclareMathAlphabet{\mathcalligra}{T1}{calligra}{m}{n}

\makeatletter
\newcommand*{\rom}[1]{\expandafter\@slowromancap\romannumeral #1@}
\makeatother

\begin{document}

\preprint{AIP/123-QED}

\title{Electrophoresis of active Janus particles}

\author{P. Bayati}
\altaffiliation[present address:]{ LPTMS, UMR 8626, CNRS, Univ. Paris-Sud, Université Paris-
Saclay, 91405 Orsay, France; E-mail:parvin.bayati@u-psud.fr}
\affiliation{Department of Physics, Institute for Advanced Studies in Basic Sciences (IASBS), Zanjan 45137-66731, Iran}%

\author{A. Najafi}%
\affiliation{Department of Physics, Institute for Advanced Studies in Basic Sciences (IASBS), Zanjan 45137-66731, Iran}%
\affiliation{Research Center for Basic Sciences \& Modern Technologies (RBST), Institute for Advanced Studies in Basic Sciences, Zanjan, Iran}%

\date{\today}

\begin{abstract}
We theoretically consider the dynamics of a self-propelled active Janus motor moving in an external electric field. The external field can manipulate the route of a Janus particle and enforce it to move towards the desired targets. 
To investigate the trajectory of this active motor, we use a perturbative scheme. At the leading orders of surface  activity of the Janus particle and also the external field, the orientational dynamics of the Janus particles behave like a mathematical pendulum with an angular 
the velocity that is sensitive to both the electric field and surface activity of the motor.
\end{abstract}

\maketitle


\section{Introduction}
Among all micro-swimmers 
\cite{happel2012low,purcell,lauga4},
active Janus particles are the novel and one of the most interesting micro-motors
with many promising and realized applications \cite{walther2008janus,walther2013janus}.
Since its first realization~\cite{paxton1}, many applications of active Janus particles 
have been achieved successfully, including effective and intelligent cargo \cite{sundararajan,baraban,gao2012cargo,yoshizumi2015trajectory,guo2018electric,sundararajan2018engineering} 
and drug delivery \cite{patra,kagan2,mou,wu,kei2014multiple,li2017micro}, the ability for entering into 
living cells \cite{gao,wangwei,wang2,shao2018erythrocyte},
detecting and healing micro-defects in microchips \cite{li2015self,zhang2018review}, and
nano-patterning techniques \cite{li2014nanomotor}.
Also it has been shown that these motors are able to clean water from organic or inorganic pollutants such as 
bacteria and heavy metals compounds more effectively
\cite{wang4,ali2015novel,soler,vilela2016graphene,zhang2017light,vilela2017microbots,wang2017internally,delezuk2017chitosan,jurado2017magnetocatalytic}. 
Designing chemical and bio-sensors is another progress in applying of active Janus particles 
\cite{kagan1,yi2016janus,li2017micro}.

Like the first synthesized models \cite{paxton1,paxton2,howse,buttinoni},
active Janus motors usually are made in the form of micron-sized rods or spheres consisting of 
two parts with distinct surface chemical properties. These motors operate by setting up 
decomposition of  chemical molecules  in the fluid
and converting chemical energy into mechanical work, often in the form of a directed or rotational motion
\cite{catchmark2005directed,qin2007rational,moran2010,popescu,ebbens2}.
Besides substantial efforts devoted to understanding the mechanism involved in the motion of a single motor
\cite{golestanian2,moran2011,walther2008janus,lauga1,Seifert,sharifi1,moran2017,katuri2016designing,
bradley2016clickable,michelin2017geometric,dong2017visible,zhou2017visible,palagi2018bioinspired},
studying their dynamics in complex environments, such as geometrical confinements and obstacles, crowded environments with high  densities of active or passive particles, and
non-Newtonian media,
is also of great interest in order to 
control their behavior in various applications \cite{volpe2011microswimmers,solovev2013collective,zottl2014hydrodynamics,
masoud2014collective,Bechinger2016_rev,di2016active,crowdy13,uspal,uspal2016guiding,zottl2016emergent,
simmchen2016topographical,das2015boundaries,liu2016bimetallic,fischer2016a,semeraro2018microstructure,lozano2018run}.
In  most of applications,  
having ability to precisely align and guide Janus motors in predefined directions towards targets is a  crucial experimental need \cite{yoshizumi2015trajectory,guo2018electric,demirors2018active}. 
Modulating the velocity of self-phoretic Janus motors by applying a gradient to the concentration of involved chemical molecules   is examined experimentally \cite{sundararajan,gao2012cargo,guo2018electric}.
In some biological applications,  a ferromagnetic core such as nickel, 
has been attached to Janus motors so that torque from  an external magnetic field can change the direction  
\cite{sundararajan,baraban,gao2012cargo,yoshizumi2015trajectory}.
Optical and chemical gradients \cite{lozano2016phototaxis,li2016light,thakur2011interaction}, 
activation field gradients \cite{bickel2014polarization,geiseler2017self}, and 
acoustic tweezers \cite{xu2015reversible}
are some other ways to guide active Janus particles.

Electrical properties of most catalytic micro-motors \cite{paxton1,moran2010,moran2017}
suggest that an electric field can be an alternative or even better candidate to control the motion of self-phoretic particles.
Moreover, pick-up and release of  cargos through induced dipolar interactions between a Janus motor and the cargos can be achieved
more easily by applying an electric field rather than other tools \cite{demirors2018active,guo2018electric}.
In this regards, the electrophoresis of catalytic Janus particles have been
investigated experimentally in Ref.~\citenum{das2015boundaries} (see Fig. 3). A spherical polystyrene colloidal particle  with a platinum cap immersed in the solution of $H_2O_2$ has been considered.
Driven by self-phoresis, the particle moves towards polystyrene side.
Their observations show that applying an electric field cause the particle to rotate and re-align its direction of symmetry parallel to the field.
Therefore, the particle moves in the direction of the electric field.

Here, we theoretically investigate  the dynamics of an  active Janus particle in the presence of a uniform external electric field.
The  particle is a non-conducting spherical colloid with  active chemical sites that are distributed asymmetrically over it. This active  particle is  immersed in an electrolyte solution. As a result of chemical reactions, a gradient of cations and anions will be developed in the media and this eventually provides 
the self-propulsion force necessary to operate the motor \cite{bayati2016dynamics}. Furthermore,  an external electric field  is also applied to this system.   

The  electric field acted on  a passive charged particle  causes the well known phenomena of electrophoresis \cite{russel, ohshima}. 
For  active colloids,  in addition to the electrophoresis, the external electric field will affect the dynamics of the Janus particle 
by modifying  and deforming the electrostatic potential and Debye layer around that particle.  
We will show that this effect has a main role in the dynamics of the active Janus particle by imposes an aligning torque on the particle.

The organization of this paper is as follows: in section \ref{sec:model}, we introduce the model of active Janus motor.
Sections \ref{sec:equations} and \ref{sec:thinDebye} are devoted to outline the main equations governing the motion of the motor 
and introduce  the approximations to simplify the equations.
We proceed by providing a perturbation expansion and obtaining analytical results in section \ref{sec:perturbation}.
Finally, the results and discussions are presented in  section \ref{sec:results}.



\section{Model system}\label{sec:model}
\begin{figure}[h!]
 \begin{center}
 \includegraphics[width=0.45\textwidth]{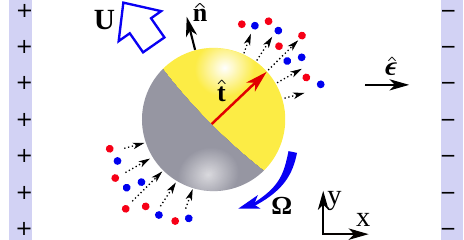}
      \caption{\label{model} 
Illustration of a spherical and chemically active particle in a uniform electric field ${\bf E}=E~\hat{\pmb \epsilon}$.
The surface of the particle consists of two hemispheres with distinct chemical activity.
Chemical reactions release  both positive and 
negative ions (blue and red dots)  on the
north hemisphere (depicted in yellow) and annihilate them on the south hemisphere 
(depicted in gray).
The unit vectors ${\bf \hat{t}}$ and $\hat{\pmb \epsilon}$
indicate the symmetry axis of the active particle  
and the direction of electric field, respectively.
Self-propulsion and the electrophoresis of the particle allow it to move 
with translational and rotational velocities denoted by
$\mathbf{U}$ and $\mathbf{\Omega}$, respectively.
   }
  \end{center}
\end{figure}

As shown in Fig.~\ref{model}, consider a spherical active particle with radius $a$ immersed in a bulk
electrolyte solution with electric permeability $\varepsilon_r$ and hydrodynamic viscosity $\eta$.
For simplicity, we consider the case of a symmetric 1:1 electrolyte, 
i.e., the valencies of two ionic species, cations and anions, are  $Z_{\pm} = \pm 1$, respectively. 
We denote the bulk number density of each ionic species by $n_{+}^b = n_{-}^b := n_{\infty}$
and their diffusion constants as $D$. 
The particle is charged and the electrostatic potential on the surface of the particle, relative to the potential in 
the bulk solution, is denoted by $\psi_{s}$. 
As demonstrated in Fig.~\ref{model},
the surface of the particle consists of two parts with distinct chemical activity~\cite{bayati2016dynamics}.
The activity of the surface of the particle triggers a set of chemical reactions asymmetrically;
both ionic species are released (``emitted'') simultaneously  on the north hemisphere of the particle 
with rate ${\dot Q}$. On the other hand, 
both cations and anions are reduced (``annihilated'') on the south hemisphere by the same rate given at emitting part. 
Physically, this can be realized, e.g., by breaking up neutral molecules (``fuel'') present in the solution
on the emitting side that drives the whole system out of chemical equilibrium. Then, on the other side of the particle, reactions start to re-combine the extra ions to neutral molecules in order to bring back the system 
to chemical equilibrium with conserved number of molecules.
Moreover, the particle is assumed to be non-conductive, therefore,
the chemical reactions on its surface do not have any electrical effect on the particle, i.e.,
releasing/annihilating of ions doe not change the uniform surface charge of the particle. Such a Janus motor is often made of 
an insulator, e.g. silica or polystyrene, with a thin layer of a catalyst coated on some part of the particle~\cite{ebbens2}.
The physical behavior  of other types of active Janus motors, such as light-activated Janus particles moving in binary mixtures~\cite{buttinoni}, 
 have some similarities to the model described here.

In addition to the model described above, other similar theoretical models for an active Janus particle 
are possible which share the same concept of releasing/annihilating of ions (or other types of solute particles) to create phoretic forces~\cite{golestanian2,brown2017ionic}. 
However, the distribution of reactions recombining of extra ions or solutes molecules differ in these models. 
Since all of these models share the same physics, and especially leads to similar results for an isolated Janus particle~\cite{bayati2019}, 
in this study, we focus only on the simple model explained above.

Following the chemical processes, local changes in the 
density of ions around  the particle will develop and subsequently drive the system out of mechanical equilibrium.
For such an isolated active Janus motor and in the absence of external field, we denote the self-propulsion active velocity by ${\bf U}^{a}$. 
From the other hand and for a passive Janus particle with a constant and symmetric surface potential,  moving in an external electric field, the phenomenon of electrophoresis  
exerts a force to the particle. We denote this electrophoretic velocity with ${\bf U}^e$.
Here, we would like to consider both mechanisms of activity and electrophoresis, simultaneously.  Decomposing the 
active part and electrophoresis parts as:
\begin{equation}
{\bf U} = {\bf U}^{a} + {\bf U}^e + {\bf U}^{ae},~~~{\bf \Omega} = {\bf \Omega}^{ae}.
\label{decompose}
\end{equation}
We aim to calculate the contributions to the linear and angular velocities  denoted by ${\bf U}^{ae}$ and ${\bf \Omega}^{ae}$   that are functions 
of the interplay between activity and electrophoresis. Promised by  symmetry principles, neither activity nor electrophoresis are able to 
solely apply torque on the particle and enforce it to rotate. This is reflected in the above equation for the  angular velocity where 
contributions from activity or electrophoresis are not considered.
Furthermore,  we will see at the next parts that the above form of decomposition will help us in developing some approximating procedures to 
deal with the complicated mathematics of this problem.

In the following, we proceed with giving the main equations governing the dynamics of the motor. 
Then, by introducing some simplifications, we will get a set of equations that can be solved analytically.


\section{Governing equations}\label{sec:equations}
Fluid velocity   ${\bf u}({\bf r})$, pressure  $p({\bf r})$, electric potential $\psi({\bf r})$ and ionic densities $n_{\pm}({\bf r})$ are main 
fields in our problem that need to be determined. Before writing the dynamical equations, it is convenient to define a dimensionless system 
of units.   We use the radius of the Janus particle $a$, thermal potential $\psi_0 = (k_B T /e)$, 
bulk density of ions $n_\infty$, and a characteristic velocity  given by
$v_0 = (k_B T /e)^2(\varepsilon_r / \eta a)$   to make 
all quantities  non-dimensional,
where $k_B$ is Boltzmann constant, $T$ is the temperature,  $\rho$ and $\eta $ are the fluid 
density and viscosity, respectively.
In this dimensionless system,  
the hydrodynamics of the incompressible Newtonian fluid is governed by the following Stokes and continuity  equations as:
\begin{equation}
\nabla ^2 {\bf u}({\bf r})-\nabla p({\bf r})+ \nabla^2 \psi({\bf r}) \nabla\psi({\bf r})=0,~~~\nabla\cdot\ {\bf u} ({\bf r})=0.
\label{stokes}
\end{equation}
One should note that in writing the above equations, we have assumed that 
 Reynolds number defined as $\mathrm{Re = (\rho U a/\eta) }$   is very small for micro-scales where  $U$ is a characteristic velocity of the motor. 
The electrostatic interactions of ions with the local potential $\psi({\bf r})$ leads to a 
distribution of body forces in the fluids given by $\nabla^2\psi({\bf r})\nabla\psi({\bf r})$. 
The electrostatic potential obeys the Poisson equation,
\begin{equation}
\delta^2 \nabla^2\psi=-\frac{1}{2}(n_{+} - n_{-})\,,
\label{poisson}
\end{equation}
where $\delta= 1/(\kappa a)$ with $1/\kappa  = \sqrt{\varepsilon_r k_BT / 2 e^2 n_\infty}$  
gives the dimensionless thickness of ``Debye layer''.
This Debye layer measures the the equilibrium thickness of 
the fluid around a colloidal particle where counter ions are accumulated there
and screen the charge of colloid~\cite{russel,ohshima}.
The distribution of the ionic species is governed by the continuity equation for the corresponding number densities,
 \begin{equation}
\frac{\partial n_\pm({\bf r})}{\partial t}+\nabla\cdot{\bf j}_{\pm}({\bf r})=0.
\label{continuty}
\end{equation}
The ionic currents ${\bf j}_\pm({\bf r})$ are given by the phenomenological expressions as:
\begin{equation}
\label{currentdensity}
{\bf j}_{\pm}({\bf r})=-\nabla n_\pm({\bf r})\mp  n_\pm({\bf r})\nabla \psi({\bf r})+
{\cal P}e\ n_{\pm}({\bf r}){\bf u}({\bf r}).
\end{equation}
Two first terms in the right hand side of above equation represent the transport by diffusion of ions and drift due to
the electric force, respectively. The third term is the convection due to the flow of fluid. Here P{\'e}clet is a dimensionless 
number that describes the ratio of the transport via convection 
by the flow to  the  thermal diffusion and it is given by ${\cal P}e = (a v_0/D)$.  

The equations \ref{stokes}, \ref{poisson}, and \ref{continuty} must be solved subject 
to appropriate boundary conditions at the surface of the motor and at infinity.
We start with the the boundary conditions at the surface of the motor. 
The chemical activity on the surface of the Janus motor is given by the following condition on the ionic fluxes:
\begin{equation}
\label{surfaceflux}
{\bf \hat{n}} \cdot {\bf j}_{\pm}({\bf r}={\hat {\bf n}}) =  {\dot q} ~{\bf \hat{t}} \cdot {\bf \hat{n}}\,,
\end{equation}
where ${\dot q}={\dot Q}a/D$, is a dimensionless number defining the strength of surface activity of Janus particle. 
Unit vector locally normal to the surface 
of the motor is denoted by  ${\bf \hat{n}}$ and angular configuration of  the Janus particle is denoted by a unit vector denoted by ${\bf \hat{t}}$. 
The electrostatic potential and the fluid velocity in a co-moving frame should satisfy
the following boundary condition at the surface of the motor:
 \begin{equation}
  {\bf u}({\bf r})=0,~~~~\psi({\bf r})=\psi_s, ~~~~~~ {\bf r} = \hat{{\bf n}}.
  \end{equation}
  At infinity, the boundary conditions read:
 \begin{equation}
  {\bf u}({\bf r})=-{\bf U}-{\bf \Omega} \times {\bf r},~~~ 
 \nabla \psi({\bf r})= - \epsilon ~ \hat{\pmb \epsilon},~~~
n_\pm= 1,~~~~ r\rightarrow \infty,
  \end{equation}
where $\epsilon= E~ ea / (k_B T) $ is a dimensionless number  characterizing the strength of external electric field and  
 $\hat{\pmb \epsilon}$ is a unite vector showing its  direction. 
${\bf U}$ and ${\bf \Omega}$, are the translational and angular velocities of the motor.
The motor experiences hydrodynamic and electrostatic forces and torques
that are given by the Stokes hydrodynamic and Maxwell electrostatic stress tensors. 
Considering  the fact that the motion of a micro-scale Janus particle takes place at low Reynolds condition,    net force and torque should vanish:
\begin{eqnarray}
{\bf F}=\oint_S &&\left\{-p\mathbb{I}+ \nabla {\bf u}+ (\nabla {\bf u})^T\right.\nonumber\\
&&\left. + \nabla \psi \nabla \psi - \frac{1}{2}\nabla \psi\cdot\nabla \psi \mathbb{I}\right\} \cdot{\bf \hat{n}}\ dS=0,
\label{force_bal}
\end{eqnarray}
\begin{eqnarray}
{\pmb \tau} = \oint_S && {\bf r}\times \left\{-p\mathbb{I}+ \nabla {\bf u}+ (\nabla {\bf u})^T\right.\nonumber\\
&&\left. + \nabla \psi \nabla \psi - \frac{1}{2}\nabla \psi\cdot\nabla \psi \mathbb{I}\right\} \cdot{\bf \hat{n}}\ dS=0.
\label{torque_bal}
\end{eqnarray}
Here, $\mathbb{I}$ stands for the  unit matrix and  superscript $T$ denoted matrix transposition.


\section{approximations}\label{sec:thinDebye}
Before solving the equations, it is instructive  to consider order of magnitude  for relevant physical parameters in a typical system. 
We will see that realistic values of parameters will allow us to introduce a couple of approximations that can simplify mathematical equations 
we have presented in previous section.
A micro-motor with $a\sim 1 ~\mu \text{m}$ moves with a  velocity about  $1 ~\mu \text{m~sec}^{-1}$
in an aqueous solution with viscosity $\eta \sim 10^{-3} ~\text{Pa~sec}$~\cite{paxton1,paxton2}. Furthermore, in an electrolyte solution,
such as $0.001$ molar solution of  $KCL$ at room temperature, the diffusion and the bulk density of ions 
are of order $D\sim 10^{-9}~\text{m}^2\text{sec}^{-1}$ and $n_\infty\sim 10^{23}~m^{-3}$, respectively~\cite{ohshima}. 
So, we can see that $\delta\sim 10^{-3}$ and ${\cal P}e\sim 10^{-1}$.
Consequently, we can make great simplifications in dynamical equations by considering the conditions of  
${\cal P}e \ll 1$ and $\delta\ll 1$, to obtain approximate equations.
To further simplification of the model, we assume that the kinetics of the decomposition/reduction is in the ``reaction limited'' regime with
${\dot q}= {\dot Q} a/D n_\infty\ll 1$~\cite{brady2011},
i.e., diffusion of ions is much faster than emission/annihilation of ions.

The main simplification is related to the concept of the Debye layer. 
For most experimentally realizable systems, as noted above, the thickness of the 
Debye layer is much smaller than the size of the particle, i.e., $\delta=1/\kappa a \ll 1$. 
At this condition, we use a macro-scale description has been developed by Yariv and coworkers~\cite{yariv2,yariv3,yariv1} and divide   
the fluid domain  into two regions:
 ``inner'' (within the Debye layer) and ``outer'' (outside of the Debye layer) parts.
 In this macro-scale description, the goal is to obtain an effective macro-scale properties of the nearly electro-neutral outer region.  
 At the following part, we will see that the effective physical properties at the bulk
 can be achieved by applying  proper boundary conditions on outer surface of the Debye layer on macro-scale fields.

The equations governing the dynamics can be solved for each regions separately and finally matched the solutions 
on the edge of the double layer~\cite{ben2002nonlinear,yariv2,yariv3,bayati2016dynamics}. 
The velocity of the Janus particle,  ${\bf U}$, which we would like to evaluate, appears in the outer region problem
as a boundary condition at infinity (in the co-moving frame).

In the inner region, because of very thin  film of fluid, the surface of the motor can be approximated as a planar wall.
The fluid is in quasi-equilibrium, the density of ions relaxes very fast to the equilibrium 
Boltzmann distribution corresponding to the local electrostatic potential $\psi$, i.e., $n_\pm = e^{\mp \psi}$.
This leads to the Poisson-Boltzmann equation for the electrostatic potential within 
the double layer satisfying the boundary condition $\psi = \psi_s$ at the 
surface of the motor. {Furthermore we consider the colloidal particles in which the surface potential is uniform 
over the surface of particle.} Solving this equation gives the the electrostatic potential in the inner problem 
then, the Stokes equation with an electrostatic body force and no-slip boundary condition at the surface of the motor
will be  solved within a lubrication approximation to obtain the fluid flow profile within the double layer.
Solution for the inner region provides the required boundary values for the outer region problem, i.e.,
the values for the potential $\psi_{D}({\bf \hat{r}})$ at the edge of the double layer, from which the so-called zeta-potential follows as:
\begin{equation}
  \label{zeta_pot}
 \zeta({\bf \hat{r}})=\psi_{s}-\psi_{D}({\bf \hat{r}}), 
\end{equation}
and the ``phoretic slip velocity'', i.e., the flow velocity ${\bf V}_{s}$ at the edge of the double layer,  is given by the 
Dukhin-Derjaguin relation \cite{prieve1} as:
\begin{equation}
\label{dukhin}
{\bf V}_{s} = \zeta \nabla_s \psi- 
4 \ln \left(\cosh \frac{\zeta}{4}\right) \nabla_s \ln N,
\end{equation}
where $\nabla_s = (\mathbb{I}-{\bf \hat{n}}{\bf \hat{n}})\cdot \nabla$
denotes the derivative along the surface (to be evaluated at the edge of the Debye 
layer).

For the outer region, regularity of  $\nabla^2 \psi$ at the limit of $\delta\rightarrow 0$ implies that the left hand side of Eq.~\ref{poisson}
vanishes at the same limit of negligible Debye screening length~\cite{yariv2}. 
Consequently, one infers that in the outer region $(n_{+} - n_{-})$ is very small (of the order of $\delta^2$ or smaller).
Then,  we  will reach to our core approximation for the outer region: 
$n_{+} \simeq n_{-}$. By applying this approximation, now we can rewrite the governing equations for the outer region. 
It is convenient to  denote all the effective macro-scale fields for the outer region by capital letters.
The hydrodynamic flow is governed by Stokes equation as:
\begin{equation}
\label{stokes_m}
\nabla ^2{\bf V}-\nabla P+\nabla^2 \Psi \nabla \Psi=0,~~~\nabla\cdot\ {\bf V}=0.
\end{equation}
The ionic densities and electrostatic potential are obtained by solving the following equations:
\begin{equation}
\label{density_m}
\nabla^2 N= 0,~~~\nabla\cdot \left(N \nabla \Psi \right)=0.
\end{equation}
The effective fields at the outer region are subjected to  boundary conditions at the  edge of Debye layer ($r \approx 1$) as:
\begin{equation}
\frac{\partial N}{\partial n}=-{\dot q}~ {\bf \hat{t}} \cdot {\bf \hat{n}},~~N\frac{\partial \Psi}{\partial n}=0,~~ 
\Psi=\psi_s-\zeta, ~~{\bf V} ={\bf V}_{s},
\end{equation}
where $\frac{\partial }{\partial n}={\bf \hat{n}}\cdot \nabla $ and 
${\bf V}_{s}$ is the phoretic slip velocity given in Eq.~\ref{dukhin}.
The boundary conditions at infinity read as:
\begin{equation}
 {\bf V}({\bf r})=-{\bf U}-{\bf \Omega} \times {\bf r},~~~ 
 \nabla \Psi({\bf r})= - \epsilon ~ \hat{\pmb \epsilon}.~~~~
\end{equation}
Finally, the effective fields satisfy the same  force and torque free conditions as given before in Eqs.~\ref{force_bal} and \ref{torque_bal}.

To demonstrate how the above equations can work, consider a very simple example that corresponds to a non-active Janus particle in the absence of electric field  (${\dot q}=\epsilon=0$). 
The results can be written as:
\begin{equation}
\label{zeroth}
 N_0 = 1, ~~ \Psi_0({\bf r})=0, ~~ \zeta_0 = \psi_s, ~~ {\bf U}_0 = {\bf \Omega}_0 = 0.
\end{equation}
This implies that in the absence of  activity and external field, the system is in the equilibrium state and the particle does not move.


\section{perturbative expansion}\label{sec:perturbation}
The assumption of thin Debye layer has simplified the governing equations, 
but the equations  are still too difficult  to be solved analytically.
In order to achieve analytical solutions and
to gain physical insight into the problem, we further restrict the scope of this study to the case when
the surface chemical activity of the motor ${\dot q}$, and 
the applied electric field $\epsilon$, are small so that the activity and applied field can be treated 
as perturbations to the equilibrium state where, the particle is not active and there is no external field. 
{We will see  later at the discussion part that for a typical active Janus particle, the strength of electric field and also the 
chemical activity  
belong to an interval that naturally can justify the validity of this perturbative scheme.}
In order to implement the perturbative expansion, we decompose the full problem of electrophoretic active Janus particle to three distinct 
auxiliary problems. As defined before in Eq.~\ref{decompose}, first problem corresponds to the propulsive motion of an active Janus particle in the 
absence of external force, we denote this problem by superscript $a$. The second problem that is denoted by superscript $e$ 
corresponds to the electrophoresis of a passive colloidal particle.
Superimposing these two problems, we will need a third contribution to recover our real problem. 
Denoting this contribution by symbol $ae$, it will collect the simultaneous effects due to both activity and electric field. Now a perturbative 
expansion can be considered for the linear and angular velocities of the Janus particle as: 
\begin{eqnarray}
&&{\bf U}^{a} =\sum_{m=1}^{\infty} {\dot q}^m~ {\bf U}^{a}_{m},~~~~~~~~
{\bf U}^{e} =\sum_{m=1}^{\infty} {\epsilon}^m~ {\bf U}^{e}_{m},\nonumber\\
&&{\bf U}^{ae} =\sum_{m,n=1}^{\infty} {\dot q}^m \epsilon^n  {\bf  U}^{ae}_{mn},~~~
{\bf \Omega}^{ae} =\sum_{m,n=1}^{\infty} {\dot q}^m \epsilon^n  {\bf \Omega}^{ae}_{mn}.\nonumber
\end{eqnarray}
Due to symmetry arguments and as it is reflected in our expansion, for first two problems of isolated Janus motor and electrophoresis, 
there is no torque to change angular orientation of the particle. We will use a similar terminology to decompose and expand all relevant fields of the problem. 
In the  following sections, we will present the leading order contributions to each problem defined here. 

\subsection{Isolated active particle}
Here we consider the case that an active Janus particle moves in an electrolyte solution without applying any external electric field.
This problem has been considered before, for completeness we succinctly present  the calculations \cite{bayati2016dynamics}. 
Applying the approximations and expansion discussed above,
the governing equations up to the first order of ${\dot q}$ are simplified as follows:
\begin{subequations}
\label{Q1E0}
\begin{equation}
\nabla ^2{\bf V}^a_1-\nabla P^a_1 = 0,~~~~~\nabla\cdot\ {\bf V}^a_1=0,
\end{equation}
\begin{equation}
\nabla^2 N^a_1= 0, ~~~ \nabla^2 \Psi^a_1 = 0,
\end{equation}
\end{subequations}
which should be solved provided the following boundary conditions for the number density of ions and 
the potential on the surface of the particle:
\begin{equation}
\frac{\partial N^a_1}{\partial n}= -{\bf \hat{t}} \cdot {\bf \hat{n}}, 
~~ \frac{\partial \Psi^a_1}{\partial n}=0,~~~
\Psi^a_1=-\zeta^a_1, ~~~~ {\bf r} =\hat{{\bf n}}.
\end{equation}
The slip velocity has a complicated dependence on the density of ions and  zeta-potential.
Perturbation expansion for these variables  takes a bit more calculations that are presented at the appendix A.
Considering the solutions in equilibrium state, Eq.~\ref{VS} reveals that the phoretic slip velocity up to the order of ${\dot q}$ reads as:
\begin{eqnarray}
\label{slip_a}
{\bf V}^a_{1,s}&&= \psi_s \nabla_{s}\Psi^a_1 - 4\ln\left(\cosh \frac{\psi_s}{4}\right)\nabla_s N^a_1,  
~~~~~ {\bf r} =\hat{{\bf n}}.~~~~~~~
\end{eqnarray}
At infinity we have:
\begin{equation}
{\bf V}^a_1({\bf r})=-{\bf U}^a_1 - {\bf \Omega}^a_1 \times {\bf r} ,~~ 
 \nabla \Psi^a_1({\bf r})=0,~~~N^a_1 = 0.
\end{equation}
Finally, the force- and torque-free conditions at the first order of ${\dot q}$ can be obtained by expanding Eqs.~\ref{force_bal} and \ref{torque_bal}.

Solving the equations for the density and potential gives the following solutions:
\begin{equation}
\label{density_a}
 N^a_1 = \frac{1}{2 r^2} ({\bf \hat{t}} \cdot {\bf \hat{r}}), ~~~~~ \Psi^a_1({\bf r})=0, ~~~~~ \zeta^a_1 = 0.
\end{equation}
Using Eqs.~\ref{slip_a} and \ref{density_a}, the slip velocity is evaluated as:
\begin{equation}
\label{slip_aa}
{\bf V}^a_{1,s}= -2\ln\left(\cosh \frac{\psi_s}{4}\right) {\bf \hat{t}} \cdot (\mathbb{I} - {\bf \hat{r}} {\bf \hat{r}}) ,
~~~~~~ {\bf r} =\hat{{\bf n}}.
\end{equation}
Having the slip velocity condition, we can proceed to evaluate the velocity of the particle.
To this end, we employ the Lorentz reciprocal theorem in hydrodynamics 
which relates two solutions of Stokes equation sharing the same geometry but with different boundary 
conditions \cite{happel2012low}.
By considering our main problem (moving an active particle with phoretic slip velocity) as one of the two problems
and noting that this particle is force- and torque-free, the Lorentz theorem gives:
\begin{equation}
{\bf U}\cdot {\bf F}_{I} + {\bf \Omega} \cdot {\pmb \tau}_{I} = -\int_{|{\bf r}|=1} {\bf V}_s\cdot {\pmb \sigma}_{I} \cdot {\bf \hat{n}}\ dS,
\end{equation}
where ${\bf F}_{I}= \int_{|{\bf r}|=1} {\pmb \sigma}_{I} \cdot {\bf \hat{n}}\ dS$ 
and ${\pmb \tau}_{I} = \int_{|{\bf r}|=1} ({\bf 
r}-{\bf r}_0) \times {\pmb \sigma}_{I} \cdot {\bf \hat{n}}\ dS$
denote the force and torque exerted by the fluid on the particle in the other problem 
with corresponding stress tensor ${\pmb \sigma}_{I}$, which can be chosen arbitrary.
We choose a sphere moving with an arbitrary translational or rotational velocity and no slip boundary
condition on its surface as problem $I$ and can easily see that
the translational and angular velocity of the spherical active Janus particle are given by~\cite{saha2014clusters}:
\begin{equation}
\label{Lorentz}
 {\bf U} = -\frac{1}{4 \pi} \int_{r=1} {\bf V}_s ~dS,~~~~{\bf \Omega} = -\frac{3}{8 \pi} \int_{r=1} {\bf r} \times {\bf V}_s ~dS. 
\end{equation}
Putting the slip velocity from Eq.~\ref{slip_aa} into the above equations and, then, calculating the integrals leads to the following results for translational and rotational velocities of the motor,
\begin{equation}
\label{velocity_Q1E0}
{\bf U}^a_1= \frac{4}{3}\ln\left(\cosh \frac{\psi_s}{4}\right) {\bf {\hat t}}, ~~~~~~~~~ {\bf \Omega}^a_1 = 0.~~
\end{equation}
Following the same procedure for higher orders of ${\dot q}$ reveals that both the translational and angular velocities vanish up to the order of ${\cal O}({\dot q}^3)$.


\subsection{Electrophoresis}
When the activity of the particle is neglected, the problem drops to the electrophoresis of
a passive charged colloidal particle in an electrolyte solution, that is one of the well-known problem in the physics of 
colloidal dispersion and has been studied in different limits analytically and numerically \cite{obrien, ohshima, russel}.
Here we are assuming that for this passive Janus particle, the surface electric potential is symmetric and  constant over the particle. 
In the thin Debye layer limit and for weak electric fields, the dynamical equations for effective fields up to the first order of $\epsilon$, i.e., ${\bf V}^e_1, \Psi^e_1, N^e_1$ satisfy the same equations as Eq.~\ref{Q1E0}, but with different boundary conditions.
The particle is passive, i.e., there is no emission/annihilation of ions on the surface of the particle.
Therefore, the boundary conditions at ${\bf r} =\hat{{\bf n}}$ are given by:
\begin{eqnarray}
\label{BC_e}
&&\frac{\partial N^e_1}{\partial n}= 0, ~~ \frac{\partial \Psi^e_1}{\partial n}=0,~~
\Psi^e_1=-\zeta^e_1,~~~~~~ {\bf r} ={\bf \hat{n}},\nonumber\\
&&{\bf V}^e_{1,s}= \psi_s \nabla_{s}\Psi^e_1 - 4\ln\left(\cosh \frac{\psi_s}{4}\right)\nabla_s N^e_1, 
~~ {\bf r} ={\bf \hat{n}},~~~~~
\end{eqnarray}
where the slip velocity on the surface of the particle up to order of 
${\cal O}(\epsilon)$ is obtained according to Eq.~\ref{VS} in appendix A.
Far from the particle, the boundary conditions read:
$$
{\bf V}^e_1({\bf r})=-{\bf U}^e_1 - {\bf \Omega}^e_1 \times {\bf r} ,~~~ 
 \nabla \Psi^e_1({\bf r})=  - \hat{\pmb \epsilon} ,~~~~
N^e_1 = 0,~~~~ r \rightarrow \infty.
$$
These equations and boundary conditions result in the following solutions:
\begin{equation}
\label{potential_Q0E1}
\Psi^e_1({\bf r})=-(r+\frac{1}{2 r^2})~ \hat{\pmb \epsilon} \cdot {\bf \hat{r}}, ~~ 
\zeta^e_1 = \frac{3}{2}~\hat{\pmb \epsilon} \cdot {\bf \hat{r}}, ~~ N^e_1({\bf r}) = 0.
\end{equation}
Then, using Eq.~\ref{BC_e}, the slip velocity is calculated as:
\begin{equation}
 {\bf V}^e_{1,s}= - \frac{3}{2} \psi_s ~ \hat{\pmb \epsilon} \cdot (\mathbb{I}-{\bf \hat{r}}{\bf \hat{r}}) , 
~~~~~~ {\bf r} ={\bf \hat{n}}.
\end{equation}
Substituting this slip velocity in to the Lorentz equation, Eq.~\ref{Lorentz}, and calculating the integrals, 
the velocity of the particle can be derived as follows:
\begin{equation}
\label{velocity_Q0E1}
{\bf U}^e_1 = \psi_s ~\hat{\pmb \epsilon}, ~~~~~~~~~~ {\bf \Omega}^e_1 = 0, 
\end{equation}
which describes the electrostatic velocity of a spherical colloidal particle in Smolokowski limit~\cite{obrien}.

The contribution of orders ${\cal O}(\epsilon ^2)$ and ${\cal O}(\epsilon ^3)$ to the velocity of the particle is zero.


\subsection{Electrophoresis of the active Janus particle}
We proceed to the order of ${\cal O}({\dot q}~\epsilon)$ that both the activity and the electric field has a simultaneous  
contribution in the dynamics of the particle.
The effective fields are given by the solutions to the
following equations:
\begin{eqnarray}
\label{Q1E1}
&&\nabla ^2{\bf V}^{ae}_{11}-\nabla P^{ae}_{11} =0, ~~~~~ \nabla\cdot\ {\bf V}^{ae}_{11}=0,\nonumber\\
 &&\nabla^2 N^{ae}_{11}= 0,~~~~~~ \nabla\cdot \left(\nabla \Psi^{ae}_{11} + N^a_1 \nabla \Psi^e_1 \right)=0.
\end{eqnarray}
In this case, the relevant boundary conditions on the surface of the particle and at infinity read as:
\begin{eqnarray*}
 &&\frac{\partial N^{ae}_{11}}{\partial n} = \frac{\partial \Psi^{ae}_{11}}{\partial n}=0,~~~\Psi^{ae}_{11}=-\zeta^{ae}_{11}, ~~~ {\bf V}^{ae}_{11} = {\bf V}^{ae}_{11,s}, ~~ {\bf r} =\hat{{\bf n}},\\ \\
 && {\bf V}^{ae}_{11}({\bf r})=-{\bf U}^{ae}_{11} - {\bf \Omega}^{ae}_{11} \times {\bf r} ,~ 
 \nabla \Psi^{ae}_{11}({\bf r})= N^{ae}_{11}=  0,  ~~  r\rightarrow \infty,
\end{eqnarray*}
respectively, where ${\bf V}^{ae}_{11,s}$ is the slip velocity up to the order of ${\cal O}({\dot q}~\epsilon)$ and is taken from Eq.~\ref{VS}.
Solving the equation for density of ions gives $ N_{11} = 0$.
After inserting $N^a_1$ and $\Psi^e_1$ from Eqs.~\ref{density_a} and \ref{potential_Q0E1} into Eq.~\ref{Q1E1}, 
we see that one needs to solve the following Poisson equation to achieve $\Psi^{ae}_{11}$,
\begin{eqnarray}
 &&\nabla^2 \Psi^{ae}_{11} =\nonumber\\
 &&  \left(\frac{1}{2 r^3}+\frac{1}{4 r^6}\right)({\bf \hat{t}} \cdot \hat{\pmb \epsilon})
 +\left(-\frac{3}{2 r^3}+\frac{3}{4 r^6}\right)(\hat{\pmb \epsilon} \cdot {\bf \hat{r}})({\bf \hat{t}} \cdot {\bf \hat{r}}).~~~~
\end{eqnarray}
The solution to this equation is obtained by evaluating the following expression~\cite{jackson1999classical},
\begin{equation}
 \Psi^{ae}_{11}({\bf r}) = \frac{1}{4 \pi} \int \frac{\nabla^2 \Psi^{ae}_{11}}{|{\bf r}-{\bf r}'|} ~ d{\bf r}'\  + \ B({\bf r}),
\end{equation}
regarding that $B({\bf r})$ satisfies the Laplace equation, i.e., $\nabla^2 B({\bf r}) = 0$ with a proper
boundary condition given by:
$$\frac{\partial B}{\partial n}|_{r=1} = \frac{\partial \Phi}{\partial n}|_{r=1},$$
and $\Phi({\bf r}) = \frac{1}{4 \pi} \int \frac{\nabla^2 \Psi^{ae}_{11}}{|{\bf r}-{\bf r}'|}~ d{\bf r}$.
A direct calculation of the above integral 
(presented in appendix B) reveals that $\Psi^{ae}_{11}$ has a form as:
\begin{eqnarray}
\Psi^{ae}_{11}({\bf r})&&= \frac{1}{4} \left(\frac{1}{3 r^3}-\frac{1}{r}\right) ~ ({\bf \hat{t}} \cdot \hat{\pmb \epsilon})\nonumber\\
&& + \frac{1}{4} \left(\frac{1}{r}-\frac{1}{r^3}+\frac{1}{2 r^4}\right) ({\bf \hat{t}} \cdot {\bf \hat{r}}) (\hat{\pmb \epsilon} \cdot {\bf \hat{r}}).
\end{eqnarray}
Then, by considering the boundary conditions on the surface of the particle,
we can calculate the changing in the zeta-potential up to ${\cal O}({\dot q}~\epsilon)$ as:
\begin{equation}
\zeta^{ae}_{11} = - \frac{1}{6} ({\bf \hat{t}} \cdot \hat{\pmb \epsilon}) + \frac{1}{8} ({\bf \hat{t}} \cdot {\bf \hat{r}}) (\hat{\pmb \epsilon} \cdot {\bf \hat{r}}).
\end{equation}
Using the density of ions and the electric potential, we evaluate 
the slip velocity up to the order of ${\dot q}~ \epsilon$ as: 
\begin{eqnarray}
{\bf V}^{ae}_{11,s} &&= \frac{1}{8} \psi_s~ {\bf \hat{t}} \cdot \left(\mathbb{I} {\bf \hat{r}} +  {\bf \hat{r}} \mathbb{I} 
- 2 {\bf \hat{r}} {\bf \hat{r}}  {\bf \hat{r}}  \right) \cdot \hat{\pmb \epsilon}\nonumber\\
  && - \frac{3}{4}\tanh \frac{\psi_s}{4}~{\bf \hat{t}} \cdot (\mathbb{I}{\bf \hat{r}}-{\bf \hat{r}}{\bf \hat{r}}{\bf \hat{r}}) \cdot \hat{\pmb \epsilon},
\end{eqnarray}
consequently, the translational and rotational velocity of the particle are attained by computing the integrals in Eq.~\ref{Lorentz} as:
\begin{eqnarray}
 {\bf U}^{ae}_{11} 
 = 0,~~~~~~~ {\bf \Omega}_{11} &&= \frac{3}{8} \tanh \frac{\psi_s}{4}~ \hat{\pmb \epsilon} \times {\bf \hat{t}}.
\end{eqnarray}
This contribution originates from the  zeta-potential modification. 
One can see that up to the order of  ${\cal O}({\dot q} \epsilon)$, the main effect of the electric field
 is to rotate it to along a direction parallel or anti-parallel to  the electric field, depending on the sign of $\psi_s$.

Carrying out the same procedure, we found that perturbation terms proportional to  ${\cal O}(\epsilon)^2$ and ${\cal O}({\dot q})^2$, are zero.
The details of these calculations are not included here but, we continue our calculations to find the next non zero contributions in our perturbation analysis.
The next  non-zero contribution is  of the order of ${\cal O}({\dot q} \epsilon^2)$, 
that we will considered its detail at the next part. 


\subsubsection*{Calculating ${\bf U}^{ae}_{12}$ and ${\bf \Omega}^{ae}_{12}$}

Here, we take into consideration the second order of electric field and try to fine the velocity of the motor up to the order ${\cal O}({\dot q} \epsilon^2)$.
To do this, the following equations should be solved,
\begin{eqnarray}
&&\nabla ^2{\bf V}^{ae}_{12}-\nabla P^{ae}_{12}+ \nabla^2 \Psi^{ae}_{11} \nabla \Psi^e_1 = 0,~~~~~
\nabla\cdot\ {\bf V}^{ae}_{12}=0,\nonumber\\
&&\nabla^2 N^{ae}_{12}= 0,~~~~~~~~~~~~\nabla\cdot \left(\nabla \Psi^{ae}_{12} + N^a_1 \nabla \Psi^{ae}_{11} \right)=0.~~~~
\end{eqnarray}
These equations are subject to the boundary conditions on the surface of the motor and are given by:
\begin{equation*}
\frac{\partial N^{ae}_{12}}{\partial n} = \frac{\partial \Psi^{ae}_{12}}{\partial n}=0,~~~
\Psi_{12}=-\zeta^{ae}_{12}, ~~~ {\bf V}^{ae}_{12} = {\bf V}^{ae}_{12,s}, ~~~ {\bf r} =\hat{{\bf n}},
\end{equation*}
and at infinity, the conditions are given by:
$$
{\bf V}^{ae}_{12}({\bf r})=-{\bf U}^{ae}_{12} - {\bf \Omega}^{ae}_{12} \times {\bf r} ,~~~~ 
 \nabla \Psi^{ae}_{12}({\bf r}) = N^{ae}_{12} = 0,  ~~~~  r\rightarrow \infty,
$$
where ${\bf V}^{ae}_{11,s}$ is taken from Eq.~\ref{VS}.
Solving these equations we derive the results for density, electric potential and zeta-potential as:
\begin{equation}
  N^{ae}_{12} = 0, ~~ \Psi^{ae}_{12}({\bf r})=0, ~~ \zeta^{ae}_{12} = 0.
\end{equation}
Although all fields in this order are zero, according to Eq.~\ref{VS}, the slip velocity has some contributions from the lower orders and is obtained as:
\begin{eqnarray}
 {\bf V}^{ae}_{12,s} &&  
 = \frac{1}{4} ~({\bf \hat{t}} \cdot \hat{\pmb \epsilon})(\mathbb{I}-{\bf \hat{r}}{\bf \hat{r}}) \cdot \hat{\pmb \epsilon}
+ \frac{3}{16} ~{\bf \hat{t}} (\hat{\pmb \epsilon} \cdot {\bf \hat{r}})^2\nonumber\\
&& - \frac{9}{64} \left(1-\tanh^2 \frac{\psi_s}{4}\right)~ {\bf \hat{t}} \cdot 
(\mathbb{I} - {\bf \hat{r}}{\bf \hat{r}}) (\hat{\pmb \epsilon}\cdot {\bf \hat{r}})^2\nonumber\\
&& - \frac{3}{16}  ~({\bf \hat{t}} \cdot {\bf \hat{r}}) (\hat{\pmb \epsilon}\cdot{\bf \hat{r}})^2 ~ {\bf \hat{r}}.
\end{eqnarray}
Finally, ${\bf U}^{ae}_{12}$ and ${\bf \Omega}^{ae}_{12}$ can be obtained by inserting the above equation into Eq.~\ref{Lorentz} and taking the integrals as:
\begin{eqnarray}
{\bf U}^{ae}_{12} &&=
\frac{1}{960} \left(-151 + 27 \tanh^2 \frac{\psi_s}{4}\right)  ~({\bf \hat{t}} \cdot \hat{\pmb \epsilon}) \hat{\pmb \epsilon}\nonumber\\
  &&- \frac{1}{64} \left(1 + 3 \tanh^2 \frac{\psi_s}{4}\right) ~{\bf \hat{t}},\\ \nonumber\\
{\bf \Omega}^{ae}_{12} && = 0.\nonumber
\end{eqnarray}
In the next section, we show in details how the trajectories of 
active Janus particle will be modified by applying an external uniform electric field.

\begin{figure}[h!]
 \begin{center}
 \includegraphics[width=0.45\textwidth]{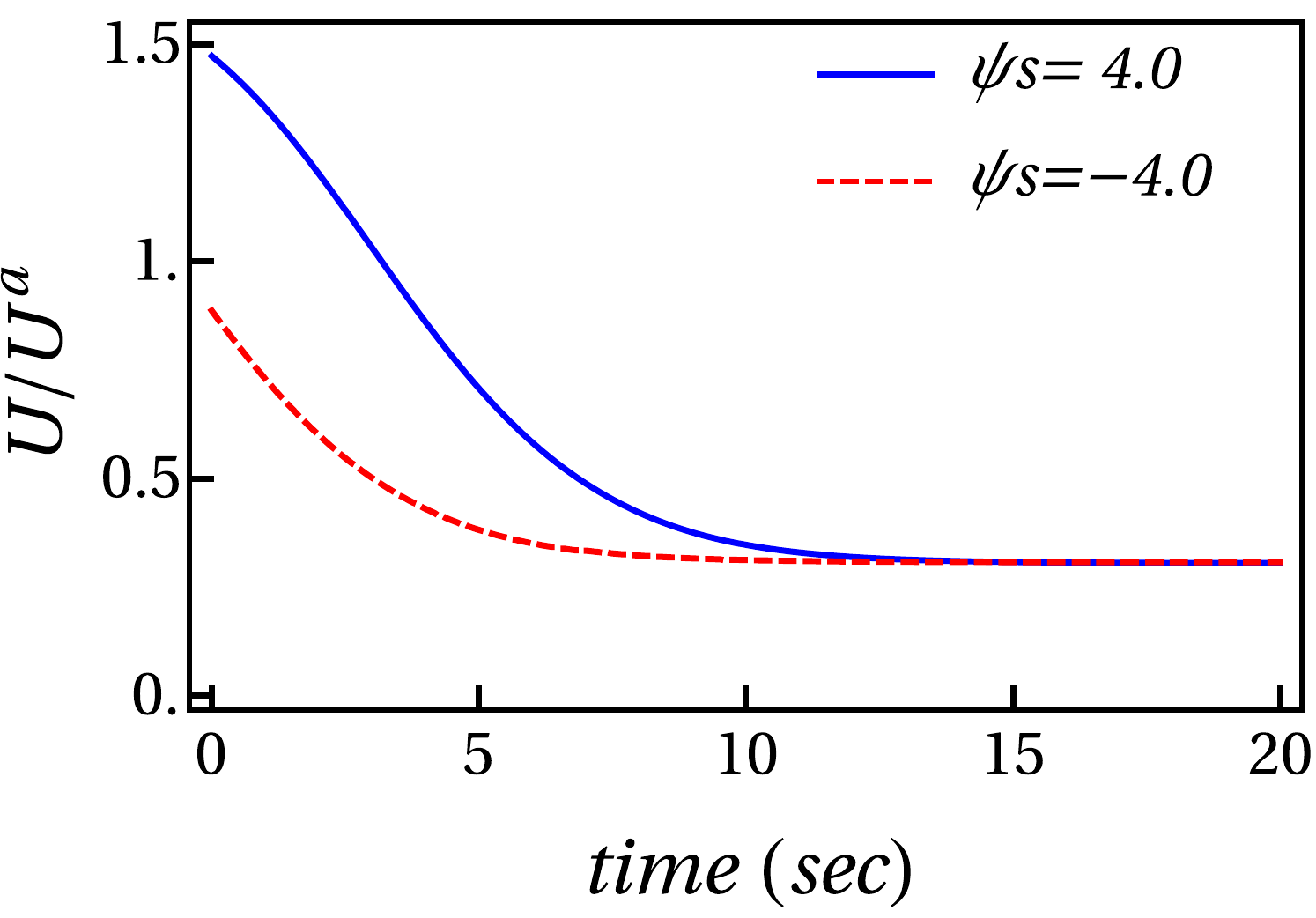}
    \caption{\label{fig2}
     The speed of the motor in the presence of an electric field normalized by its intrinsic speed $U^a$
     as a function of time. This plot corresponds to initial configuration that
     $\theta = \pi/3$ for two values of surface 
     potential and $\epsilon =0.01$ and ${\dot q}=0.1$.
   }
  \end{center}
\end{figure}

\section{results and discussion}\label{sec:results}
Combining all the above results, the total translational and rotational velocity of the motor up to the leading orders of perturbation analysis
are given by:
\begin{eqnarray}
\label{U}
 {\bf U} &&= \frac{4}{3}\ln\left(\cosh \frac{\psi_s}{4}\right) {\dot q} ~{\bf {\hat t}} + \psi_s \epsilon ~\hat{\pmb \epsilon}\nonumber\\
 &&+\frac{1}{960} \left(-151 + 27 \tanh^2 \frac{\psi_s}{4}\right) {\dot q}~\epsilon^2~({\bf \hat{t}} \cdot \hat{\pmb \epsilon})\hat{\pmb \epsilon}\nonumber\\
  &&- \frac{1}{64} \left(1 + 3 \tanh^2 \frac{\psi_s}{4}\right) {\dot q}~\epsilon^2 ~{\bf \hat{t}},
\end{eqnarray}
\begin{equation}
\label{omega}
 {\bf \Omega} = \frac{3}{8} \tanh \frac{\psi_s}{4}{\dot q}~\epsilon ~ \hat{\pmb \epsilon} \times {\bf \hat{t}}.
\end{equation}
One should note that the above equations are written in dimensionless units. 
As noted before, the linear and angular speeds given by: $v_0$ and
$v_0/a$ with $v_0 = (k_B T /e)^2(\varepsilon_r / \eta a)$, can be used  to recover the physical dimensions.

{ In typical systems of active Janus motors and electrophoresis experiments \cite{paxton1,das2015boundaries},
one has
$n_\infty \sim  10^{23} ~\text{m}^{-3}, {\dot Q} \sim 10^7~\text{sec}^{-1}\mu m^{-2}, \psi_s \sim 100~\text{mV}$ and $E \sim 250~\text{V m}^{-1}$,
which are equivalent to dimensionless values as ${\dot q} \sim 0.1, \psi_s \sim 4.0$ and $\epsilon \sim 0.01$, regarding that at the room temperature we have $k_B T/e \sim 25~\text{mV}$.
One should note that, for such typical swimmer that is  realizable in experiments, we have $\epsilon\ll 1$ and ${\dot q}\ll 1$ that justifies the 
validity and convergence of the perturbative expansion we have used in our analysis. 
Using these values, one can estimate the magnitude of the translational and the angular velocities of the motor as $U \sim 60 ~\mu\text{m} /s$ and $\Omega \sim 0.2~\text{sec}^{-1}$, respectively.}

Interestingly, we note that,
as we neglected thermal fluctuations in this study, 
 motion of the motor is constrained to take place  in a two dimensional plane
containing  two vectors ${\bf \hat{t}}(0)$ (initial orientation of the swimmer) and $\hat{\pmb \epsilon}$, direction of the electric field.
For simplicity, we choose the $x$ axis as the direction of electric field $\hat{\pmb \epsilon}$, in this case ${\bf \hat{t}}$ lies in the $x-y$ plane such that $\cos \theta = {\bf \hat{t}} \cdot \hat{\pmb \epsilon}$.
In the following, we consider the case that ${\dot q}=0.1, \epsilon = 0.01$.
We start presenting results with investigating the effluence of the external electric field on the speed of the motor.

First we discuss the case of positive surface potentials, i.e., $\psi_s > 0$.
From Eq.~\ref{U}, one infers that applying an electric field in the direction of ${\bf \hat{t}}$, i.e, $\theta=0$
causes the speed of the motor to increase with respect to the speed of an isolated motor.  
According to Eq.~\ref{omega}, this case does not induce any rotational motion for the swimmer. 
This increasing of the speed can be understood by considering the fact that, in this case both electrophoresis and activity lead the motor
to move in the same direction, and thus both effects cooperate in propelling the motor in the direction of electric field with an enhanced speed.
For the electric field applied in the direction of $-{\bf \hat{t}}$, i.e, $\theta=\pi$,
 two contributions drive the motor in opposite directions, therefore, 
the speed  decreases with respect to the speed of the motor in the absence of the field. 

On the other hand, the  case of a motor with negative surface potentials is different.
For $\psi_s <0$, the speed of the motor decreases for $\theta=0$ and increases for $\theta=\pi$.
For other initial orientations of the motor, the speed decreases due to applying the electric field
to reach the same value irrespective of the sign of surface potential. 
Fig.~\ref{fig2}, illustrates results for two typical examples. Here the surface potential is chosen as  $\psi_s = \pm 4.0$, 
initial orientation is given by  $\theta= \pi/3$ and other physical  values are ${\dot q}=0.1, \epsilon = 0.01$.  As one can see, in both cases 
proceeding in time the speed will decrease.

The most interesting impact of the electric field is its influence on  orientation and direction of the Janus particle.
In order to find out how the electric field affects the dynamical behavior of the motor, we
follow some trajectories which start at the same position but with different initial orientations of the motor
with respect to the direction of the electric field. 
The dynamics of the active particle is obtained by the following equations:
\begin{equation}
 \label{eq:dynamics}
 \frac{dx}{dt} = U_x,~~~~~~~~~~
 \frac{dy}{dt} =  U_y,~~~~~~~~~~
 \frac{d\theta}{dt} =  \Omega =  \omega \sin \theta,
\end{equation}
where all variables are dimensionless, $\omega = \frac{3}{8} \tanh \frac{\psi_s}{4}{\dot q}~\epsilon$ and $U_x$, $U_y$ and $\Omega$ are given in equations \ref{U} and \ref{omega}.
Trajectory of the motor in the presence of an electric field is obtained by integrating the above equations.

\begin{widetext}

 \begin{figure}[h!]
 \begin{center}
 \includegraphics[width=0.8\textwidth]{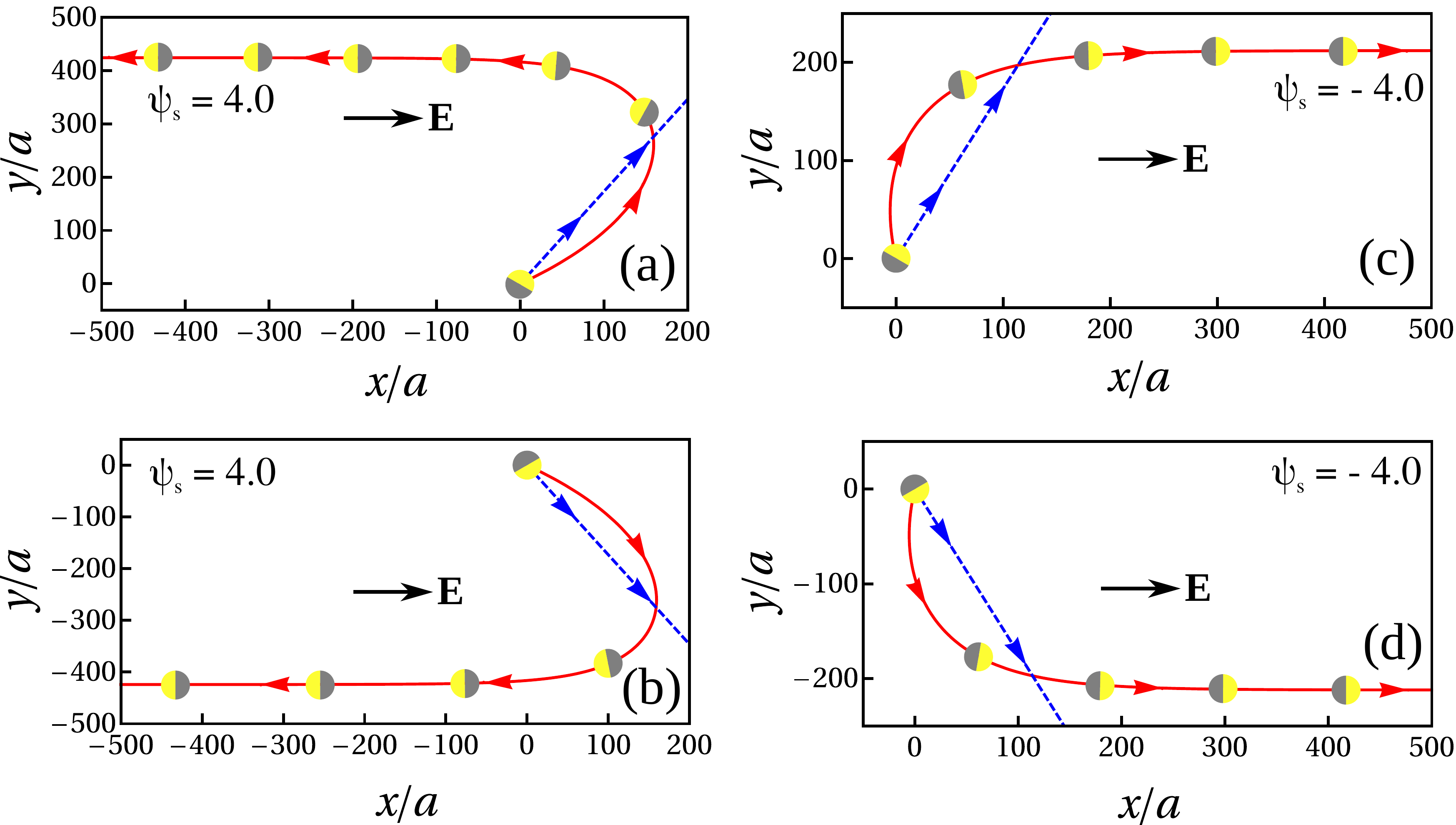}
    \caption{\label{fig3} 
    Trajectories of the Janus motor in the presence of an external uniform electric field. The panels correspond to typical values ${\dot q}=0.1, \epsilon = 0.01$ and 
    to different surface potential
    (a,b) $\psi_s=4.0$ and (c,d) $\psi_s=-4.0$ for initial orientations of the motor
    respect to the direction of the electric field (a,c) $\theta_i = \pi/3$ and (b,d) $\theta_i = -\pi/3$. The dashed blue lines show the trajectory of a corresponding isolated motor, i.e., in the absence of electric field. 
    The colored small disks indicate the motor orientation along its trajectory.
    The emitting side of the motor is depicted by yellow, as
    shown in Fig.~\ref{model}.
        }
  \end{center}
\end{figure}

\end{widetext}

For general cases, Eq. (44) can be solved numerically to give the trajectory. Some typical examples of the trajectories are illustrated in Fig.~\ref{fig3} for two different initial orientations $\theta_i = \pm \pi/3$ and for two different surface potentials $\psi_s=\pm 4.0$.
As it is reflected from Eq.~\ref{omega} and also seen in Fig.~\ref{fig3}, for any initial orientation,
the electric field enforces the motor to rotate  until its orientation ${\bf \hat{t}}$ 
points parallel or anti-parallel to the direction of $\hat{\pmb \epsilon}$,  
for  cases of $\psi_s<0$ or $\psi_s>0$, respectively.
Afterwards, the motor keeps moving toward its emitting side, i.e., in the direction of the electric field for $\psi_s < 0$ and opposite to the field 
for $\psi_s > 0$. This is in contrast to electrophoresis of a passive particle. The case that the motion of a particle is in the direction of the electric field for positive surface potentials and is against the field for negative surface potentials

Similar behavior have been observed experimentally in Ref.~\citenum{das2015boundaries}
for dynamics of a half-coated Janus sphere composed of polystyrene and platinum moving in solution of $H_2O_2$. 
Their results reveal that the surface potential is negative, then the motor
reorients the polystyrene side in the direction of the applied field and 
moves towards its intrinsic direction of motion.
One should note that in that experiment the isolated Janus particle self-propels with the polystyrene side forward.

The instantaneous orientation of  the Janus particle is shown by colored disks  along its trajectory  in Fig.~\ref{fig3}. 
As illustrated is  Fig.~\ref{model}, here we have shown the  emitting side with yellow colors.  In addition, the time intervals between 
the symbols are the same then, 
the position of disks  are an indication of  the speed of the motor. 
The distance traveled by the motor between two symbols decreases  during its motion, this implies that the speed of motor  decreases as time increases.

Although, the angular rotation tends to orient the motor to the parallel/anti-parallel direction with respect to the electric field,  Eqs.~\ref{U} and \ref{omega} reveals that the time takes the motor to re-orient diverges to infinity.
However, this is not physically insightful. In real situations, fluctuations with  thermal or non-thermal sources do not allow us to define a 
precise value for angle. We denote this unavoidable error by $\Delta\theta$
then, the final orientation of the motor is not exactly parallel/anti-parallel to the field.
Now, by considering $\psi_s >0$ and integrating the equation for $\theta$ in Eq.~\ref{eq:dynamics}, we obtain the ``relaxation time'' $\tau$, the time that takes the motor to reach its final orientation $\theta_f = \pi -\Delta \theta$ from an initial orientation 
$\theta_i$ as follows:
\begin{equation}
 \frac{\tau}{t_0}  = \frac{1}{\omega} \ln \left(\cot \frac{\theta_i}{2} \cot\frac{\Delta \theta}{2}\right),
\end{equation}
where $t_0=a/v_0 \sim 10^{-3} ~\text{sec}$ is the time scale. 
For example, for the motor considered above, i.e., for values ${\dot q}=0.1, \epsilon = 0.01, \psi_s = 4.0$ with initial orientation $\theta_i = \pi/3$ 
and for $\Delta\theta\sim1^\circ = \pi/180 ~\text{rad}$,
the relaxation time is about $10 ~\text{sec}$.
Increasing the magnitude of the electric field will decrease this time scale.
This time scale is comparable with the relaxation time that are achieved in experiment 
in which optical forces where used to reorient the particle~\cite{lozano2016phototaxis}.

\RR{As an important remark, we investigate the stability of the final state of the particle motion.
 We consider the case of small deviations from the final orientation of the particle for the case $\psi_s < 0$, i.e., $\delta\theta \ll 1$. One can do the same procedure straightforwardly for negative surface potentials.
 Using Eq.~\ref{U} we can write Eq.~\ref{eq:dynamics} as:
\begin{eqnarray}
 &&\frac{d}{dt}x(t)=(a+b)\cos\delta\theta,~~~\frac{d}{dt}y(t)=a\sin\delta\theta,\nonumber\\
 &&\frac{d}{dt}\delta\theta (t)=-\omega_0\sin\delta\theta,
\end{eqnarray}
where $a=\frac{4}{3}\ln\left(\cosh \frac{\psi_s}{4}\right) {\dot q}- \frac{1}{64} \left(1 + 3 \tanh^2 \frac{\psi_s}{4}\right) {\dot q}~\epsilon^2$, 
$b=\psi_s \epsilon+\frac{1}{960} \left(-151 + 27 \tanh^2 \frac{\psi_s}{4}\right) {\dot q}~\epsilon^2~({\bf \hat{t}} \cdot \hat{\pmb \epsilon})$ and $\omega_0= \frac{3}{8} \tanh \frac{|\psi_s|}{4}{\dot q}~\epsilon$. 
For $\delta\theta \ll 1$ we have:
\begin{equation}
\frac{d}{dt}x(t)=(a+b),~~~\frac{d}{dt}x(t)=a\delta\theta,~~~~\frac{d}{dt}\theta (t)=-\omega_0\delta\theta.
\end{equation}
For initial state given by $x=0,~y=y_0,~\delta\theta=\theta_0$ we will see that the $x$ coordinate increases linearly with time and 
\begin{equation}
\delta\theta \sim \theta_0 e^{-\omega_0t},~~~y-y_0\sim -\frac{a\theta_0}{\omega_0} (e^{-\omega_0t}-1).
\end{equation}
Therefore, at long times, the deviation from $\theta=0$ decays and the coordinate perpendicular to the electric field, i.e. $y$, have a constant value $\delta y_{\infty}\sim \frac{a\theta_0}{\omega_0}$. This means that
the final state moving along the electric field is stable against small fluctuations.
}

In this article, we have assumed that the surface potential over a passive Janus particle is uniform.  
It should be mentioned that for a passive particle, surface potential and subsequently zeta potential can have non-uniform 
distribution with different origins with respect to what we have considered here. This non-uniformity can originated from 
geometrical and electrical effects~\cite{fair1989electrophoresis}. 
For such passive particles, dipole and quadrupole distribution of zeta potential, will have extra contributions in the particle’s motion. For such particles, the dipolar and quadrupolar contributions should be included to what we have presented here.

In summary, we have theoretically studied the dynamics of a  self-propelled Janus particle in the field of an external force. 
In our description, the strength of surface activity ${\dot q}$ and the external electric field
$\epsilon$ are considered as two independent parameters and it is shown that the resulted  trajectory of the Janus particle 
is very sensitive to these parameters.  We have shown, by tuning the electric field and the surface activity, it is possible  to control the 
dynamics of the particle and operate it  in a desired manner.

{ Finally, it should be mentioned that for realistic experimental situations, interaction with confining boundaries are unavoidable. 
To have a complete control on the trajectory of active Janus particle, the effects of walls should be considered carefully. 
 }

\section{acknowledgement}
Financial support from Iran National Science Foundation (INSF), is acknowledged. 


\appendix
\label{appendix:VS}
\section{Expansion of ${\bf V}_s$ in terms of ${\dot q}$ and $\epsilon$}
Here, we present the expansion of the phoretic slip velocity:
\begin{eqnarray}
&&{\bf V}_{s} =\zeta \nabla_{s}\Psi - 4\ln\left(\cosh \frac{\zeta}{4}\right)\nabla_s N,\nonumber
\end{eqnarray}
 in terms of  ${\dot q}$ and $\epsilon$ as follows:
\begin{eqnarray}
\label{VS}
&&{\bf V}_{s}\approx \zeta_{0} \nabla_{S}\Psi_{0} -4 z_0\nabla_sN_{0}\nonumber\\
&&+{\dot q} \left(\zeta_{0}\nabla_{S}\Psi^a_{1}+\zeta^a_{1}\nabla_{S}\Psi_{0}
-4 z_0\nabla_sN^a_1-\zeta^a_{1}\tanh\frac{\zeta_0}{4}\nabla_sN_0\right)\nonumber\\
&&+\epsilon \left(\zeta_{0}\nabla_{S}\Psi^e_{1}+\zeta^e_{1}\nabla_{S}\Psi_{0}
-4 z_0\nabla_sN^e_1-\zeta^e_{1}\tanh\frac{\zeta_0}{4}\nabla_sN_0\right)\nonumber\\
&&+{\dot q}\epsilon \left(\frac{ }{ }\zeta_{0}\nabla_{S}\Psi^{ae}_{11}+\zeta^{ae}_{11}\nabla_{S}\Psi_{0}
-4 z_0\nabla_sN^{ae}_{11} \right.\nonumber\\
&&\left.
-\zeta^{ae}_{11}\tanh\frac{\zeta_0}{4}\nabla_s N_0 -\frac{1}{4}{\zeta^a_1}{\zeta^e_1}~ \text{sech}^2 \frac{\zeta_0}{4}\nabla_s N_0
\right)\nonumber\\
&&+{\dot q}\epsilon^2 \left(\frac{ }{ }\zeta_{0}\nabla_{S}\Psi^{ae}_{12}+\zeta^{ae}_{12}\nabla_{S}\Psi_{0}
-4 z_0\nabla_sN^{ae}_{12}\right.\nonumber\\
&&\left.-(\zeta^{ae}_{12}-\frac{3}{32}\zeta^a_1{\zeta^e_1}^2)\tanh\frac{\zeta_0}{4}\nabla_s N_0\right.\nonumber\\
&&\left.-\frac{1}{4}({\zeta^a_1}{\zeta^e_2}+{\zeta^e_1}\zeta^{ae}_{11})~ \text{sech}^2 \frac{\zeta_0}{4}\nabla_s N_0
\right),
\end{eqnarray}
where $z_0 = \ln \left(\cosh (\zeta_{0}/4)\right)$ and we keep only terms that lead to non-zero velocity. Higher orders can be evaluated 
straightforwardly.


\label{appendix:psi_11}
\section{Calculating $\Psi^{ae}_{11}$}
In this appendix, we calculate $\Psi^{ae}_{11}$ where
\begin{equation*}
 \Psi^{ae}_{11} = \Phi({\bf r}) + F({\bf r}),~~\nabla^2 F({\bf r})=0,~~
\frac{\partial F({\bf r})}{\partial r}|_{r=1} = -\frac{\partial \Phi({\bf r})}{\partial r}|_{r=1},
\end{equation*}
and 
\begin{eqnarray}
 &&\Phi({\bf r})  = \frac{1}{4 \pi} \int \frac{\nabla N^a_1({\bf r}') \cdot \nabla \Psi^e_1({\bf r}')}{|{\bf r}-{\bf r}'|} d{\bf r},
 \nonumber
  \end{eqnarray}
 where $F({\bf r})$ is introduced in order to satisfy the boundary condition on the surface of the motor, 
i.e., $\frac{\partial \Psi^{ae}_{11}}{\partial r}|_{r=1} = 0$, be achieved.
By plugging Eqs.~\ref{density_a} and \ref{potential_Q0E1} into the above equation, and expanding $1/|{\bf r} - {\bf r}'|$,
${\bf \hat{r}}$ and ${\bf \hat{r}'}$ in terms of spherical harmonics $Y_{nm} (\theta, \phi)$~\cite{arfken}, one obtains $\Phi$ as follows
\begin{eqnarray}
 \Phi({\bf r})  &&= \epsilon~{\dot q}~(\hat{\pmb \epsilon} \cdot {\bf \hat{t}}) 
  \left(\frac{1}{10 r^3}-\frac{1}{4r}\right)\nonumber\\
    &&+ \frac{1}{5} \epsilon~{\dot q}~({\bf \hat{t}} \cdot \hat{\bf r}) (\hat{\pmb \epsilon} \cdot \hat{\bf r})
  \left(\frac{5}{8r^4}+\frac{5}{4r}-\frac{3}{2r^3}\right).
  \end{eqnarray}
Then,  the boundary condition of $F({\bf r})$ on the surface of the motors is evaluated as
\[
\frac{\partial F({\bf r})}{\partial r}|_{r=1} = \frac{1}{20}~ (\hat{\pmb \epsilon} \cdot {\bf \hat{t}}) 
-\frac{3}{20} ~({\bf \hat{t}} \cdot \hat{\bf r}) (\hat{\pmb \epsilon} \cdot \hat{\bf r}).
\]
Solving $\nabla^2 F({\bf r})=0$ using the spherical harmonics and employing the above condition, one arrives at
\begin{equation}
 F({\bf r}) 
= -\frac{1}{60 r^3}~ {\bf \hat{t}} \cdot \left(\mathbf{I}-3 \hat{\bf r} \hat{\bf r}\right) \cdot \hat{\pmb \epsilon}.
\end{equation}
Therefore, $\Psi^{ae}_{11}$ is obtained as
\begin{eqnarray}
 \Psi^{ae}_{11} &&= \epsilon~{\dot q}~(\hat{\pmb \epsilon} \cdot {\bf \hat{t}}) 
  \left(\frac{1}{12 r^3}-\frac{1}{4r}\right)\nonumber\\
 &&+  \epsilon~{\dot q}~({\bf \hat{t}} \cdot \hat{\bf r}) (\hat{\pmb \epsilon} \cdot \hat{\bf r})
  \left(\frac{1}{8r^4}+\frac{1}{4r}-\frac{1}{4r^3}\right).
\end{eqnarray}


%

%

\end{document}